\documentclass[preprint,11pt,aps,pra]{revtex4-1}
\usepackage{graphicx}
\usepackage{amsfonts,amsmath,amssymb}
\usepackage{enumitem}
\newcommand{\e}{\text{e}}

\newcommand{\ket}[1]{|#1 \rangle}

\newcommand{\ie}{\emph{i.e.}}
\newcommand{\eg}{\emph{e.g.}}
\begin{document}


\title{Geometric Phase of a Spin-1/2 Particle Coupled to a Quantum Vector Operator}

\author{Pedro Aguilar}
\email{pedro.aguilar@nucleares.unam.mx}
\affiliation{Instituto de Ciencias Nucleares, Universidad Nacional Aut\'onoma de M\'exico\\
	PO Box 70-543, 04510, D.F., M\'exico.\\
}

\author{Chryssomalis Chryssomalakos}
\email{chryss@nucleares.unam.mx}
\affiliation{Instituto de Ciencias Nucleares, Universidad Nacional Aut\'onoma de M\'exico\\
	PO Box 70-543, 04510, D.F., M\'exico.\\
}

\author{Edgar Guzm\'an}
\email{edgar.guzman@correo.nucleares.unam.mx}
\affiliation{Instituto de Ciencias Nucleares, Universidad Nacional Aut\'onoma de M\'exico\\
	PO Box 70-543, 04510, D.F., M\'exico.\\
}

\begin{abstract}
	\noindent We calculate Berry's phase when the driving field, to which a spin-1/2 is coupled adiabatically, rather than the familiar classical magnetic field, 
	is a quantum vector operator, of noncommuting, in general, components, \emph{e.g.}, the angular momentum of another particle, or another spin. The geometric phase of the entire system, spin plus ``quantum driving field'', is first computed, and is
	then subdivided into the two subsystems, using the Schmidt decomposition of the total
	wave function --- the resulting expression shows a marked, purely quantum effect, involving the commutator of the field components.
	We also compute the corresponding mean ``classical'' phase, involving a precessing magnetic field in the presence of noise, up to terms quadratic in the noise amplitude --- the results are shown to be in excellent agreement with numerical simulations in the literature. Subtleties in the relation between the quantum and classical case are pointed out, while three concrete examples illustrate the scope and internal consistency of our treatment.
	
	\vspace{0.2cm}
	

    \vspace{0.2cm}

	\noindent {03.65.Vf, 03.65.-w}
\end{abstract}

\maketitle

\section{Introduction}
\noindent The standard setup in dealing with Berry's phase involves a hamiltonian $H(\xi)$ that
depends on external parameters $\xi^i$. A cyclic change of the latter, at a rate that
permits invoking adiabaticity, implies that a quantum state that initially coincides
with a non-degenerate eigenstate of $H$, will end up, at the end of the cycle, returning
to its initial form, picking up at most a phase. Part of this phase is the expected
dynamical one, of the energy-times-time type, but there is, in general, another
part, which depends only on the curve traced out in $\xi$-space, and not on its time-
parametrization, and is therefore called geometric~\cite{Berry08031984}.
The archetypical system exhibiting geometric phases is a spin 1/2 coupled
to a magnetic field $\mathbf{B}(t)$, the components of which play the role of the external
parameters $\xi^i$. The latter vary cyclically with time as $\mathbf{B}(t)$ traces out a closed curve (typically a circle) in $\mathbb{R}^3$. Realistically however, fields do not follow smooth curves,
rather they jiggle around average paths, due to omnipresent noise, a fact most
prominently affecting applications of geometric phases in quantum computing \cite{292c4cfc0e4847d9aeb6240fd0d707b8,PhysRevA.76.044303,Zanardi199994}.
Such considerations have spurred a marked interest in realistic setups  --- see, \eg, 
Refs. \cite{PhysRevA.39.3228,PhysRevLett.60.2339,Uhlmann1986229,%
PhysRevLett.90.190402,PhysRevLett.90.160402}. 
In particular, the authors of Ref. \cite{PhysRevLett.91.090404} studied the effect of noise, 
superimposed to the standard precessing magnetic field, and found that, to first order in the noise amplitude, the geometric phase accumulated receives no correction.

Our primary aim in this work is to take the next logical step in this direction by studying 
 the effect of the \emph{quantum} nature of the driving field to the geometric phase accumulated
by the spin \textemdash{} after all, the entire experimental setup used, including the parts
associated with the driving field, ought to be described by quantum mechanics. Since corrections to the geometric phase are expected to be quadratic in the noise amplitude, comparison with the classical result presupposes carrying out the calculation of Ref. \cite{PhysRevLett.91.090404} up to second order. Then, for the quantum case, we assume the spin is coupled to a quantum vector operator  $A$, the
components of which may well fail to commute among themselves. This operator
is associated to some quantum system, \emph{e.g.}, it could be the angular momentum of
a quantum particle, and its expectation value is taken to precess around an axis,
just like $\mathbf B(t)$ does in the familiar ``classical'' example. We compute the geometric
phase that corresponds to the spin using the Schmidt decomposition of the total
wave function \textemdash{} our main result is Eq.~(\ref{quantumPhase}). We find, as expected, that the main
contribution to the phase corresponds to the precession of the expectation value of
$\mathbf{A}$. We also find a quantum correction to this result, having to do with the quantum
fluctuations of $\mathbf{A}$ around its expected value \textemdash{} part of this correction is the not-so-obvious analogue of the classical noisy result, but part of it is purely quantum, as it involves the commutator of certain components of $\mathbf{A}$.

Spin-1/2 systems have of course been thoroughly studied, in particular  in the context of geometric phases, so much so that any new 
contribution in the field must be thoroughly justified, with a particularly heavy burden of proof regarding novelty. We explain here in what exact sense our work differs from previous ones, and why we believe it is interesting enough to warrant one more ``spin-1/2 article''.
As mentioned above, the setup considered in Ref. \cite{PhysRevLett.91.090404}  served as our starting point: our idea was to substitute quantum fluctuations for classical stochastic noise, and see if any surprises come up --- it turns out that they do. A fluctuating (stochastic) magnetic field has also been considered previously,\cite{Wan:09} the emphasis there being on the induced decoherence. The abstract of Ref. \cite{PhysRevLett.89.220404}, on the other hand,  begins like this: ``We calculate the Berry phase of a spin-1/2 particle in a magnetic field considering the quantum nature of the field'', which sounds alarmingly familiar. Still, the system considered there consists of a spin-1/2 coupled to one or two Fourier modes of a quantized magnetic field, described in principle by QED, each mode being  represented algebraically as a harmonic oscillator, and the emphasis is on vacuum-induced phases. A more complete treatment, that would allow recovering a classical-like precessing magnetic field would make possible comparisons with our results, but in their given form, the two treatments (ours and that of Ref. \cite{PhysRevLett.89.220404}) are not directly comparable.

Another problem  previously studied in the literature is the environmentally induced decoherence \cite{deChiara:2008,Wang_2004,PhysRevLett.90.160402,PhysRevLett.90.190402}, where apart from the standard external parameters, a bosonic field is also present. Other treatments involve a semi-classical approach,  
using  the Born-Oppenheimer~\cite{Mead_1992},  or the WKB~\cite{Wilkinson_1984} approximation. In Ref. \cite{PhysRevA.62.022109} the problem of subdividing the total geometric phase into spin-1/2 subsystems is tackled, but, importantly,  the spins there are not thought of as driving parameters for each other and the emphasis is on the entanglement dependence of the total geometric phase acquired by the system. Finally, in Ref. \cite{PhysRevLett.92.150406} a pair of coupled spins is considered, with one of them driven by a magnetic field. Apart from the non-standard coupling between the spins assumed,  the main object of study there  is the dependence of the geometric phase on the coupling constant, and its subdivision to the individual spins, both questions being peripheral in our work.

The structure of the paper is as follows: in section \ref{stochasticField} we carry out the
second order classical calculation, extending previous results.\cite{PhysRevLett.91.090404}
In section \ref{quantumField} we treat the quantum case, which is our main subject, and comment on its subtle relation to the classical one. We also illustrate our result with some examples. The final section contains some remarks, and points to directions for future work.
\section{Spin 1/2 coupled to a magnetic field with noise}
\label{stochasticField}
\subsection{Second order corrections to the geometric phase for a classical stochastic field}
\label{Socttgp}
\noindent Consider the hamiltonian $H$ that represents the interaction of a spin-1/2 particle with a time dependent magnetic field $\mathbf{B}(t)$,
\begin{equation}
H=\frac{\lambda}{2}\mathbf{B}(t)\cdot\boldsymbol\sigma\,,
\end{equation}
where $\lambda$ is a coupling constant and $\sigma_i$ the Pauli matrices.  If the spin state is initially aligned with the magnetic field, and the evolution of $\mathbf{B}(t)$ is periodic and adiabatic,  the geometric phase accumulated during one cycle is minus one half  the solid angle subtended by the curve $C$ traced by the direction of $\mathbf{B}(t)$  on the unit 2-sphere~\cite{Nakahara}. In particular, if $\mathbf B(t)$ precesses around the $\hat{z}$ axis, the geometric phase after a full turn is
\begin{equation}
\label{gamma0}
\varphi_{+}=-\pi(1-\cos\Theta_0)
\,,
\end{equation}
where  $\Theta_0$ is the fixed angle between the $\hat{z}$ axis and $\mathbf{B}(t)$. We now write the previous result in a form that will be useful later on. Note that  $\cos\Theta_0=\langle\sigma_z\rangle$, where  angular brackets denote the expectation value of an operator in the state of the spin. Substituting this in the expression for $\varphi_+$ we have
\begin{equation}
\label{gamma0sigmaz}
\varphi_+=-\pi(1-\langle\sigma_z\rangle)
\,.
\end{equation}
A proposal for treating stochastic fluctuations of the driving field has been carried out in \cite{PhysRevLett.91.090404} by superposing noise to the precessing field --- a perturbative analysis found zero first-order corrections to the geometric phase. We  extend, in what follows, these results to second order, and find a nonvanishing correction. 

We start by assuming the total magnetic field to be given by
\begin{equation}
\label{totalStochasticMagneticField}
 \mathbf{B}(t) = \mathbf{B}_0(t) + \epsilon\mathbf{b}(t),
\end{equation}
where $\mathbf{B}_0(t)$ precesses around the $z$-axis, and $\mathbf{b}(t)$ is a stochastic process modelling noise, the smallness of the latter being carried by the dimensionless parameter $\epsilon \ll 1$. The direction of $\mathbf{B}(t)$ is defined by polar angles $\Theta$ and $\Phi$, while the precessing $\mathbf{B}_0(t)$, assumed of unit strength,  is given by $(\sin\Theta_0\cos\omega t,\sin\Theta_0\sin\omega t,\cos\Theta_0)$ in a cartesian frame $(\hat{\mathbf{x}},\hat{\mathbf{y}},\hat{\mathbf{z}})$. 
We also use the following frame, adapted to $\mathbf{B}_0(t)$,
\begin{equation}\label{adaptedFrame}
 \begin{aligned}
  \textbf{e}_1 &  = (\cos\Theta_0\cos\omega t,\cos\Theta_0\sin\omega t,-\sin\Theta_0) \\
  \textbf{e}_2 &  = (-\sin\omega t,\cos\omega t,0), \\
  \textbf{e}_3 &  = (\sin\Theta_0\cos\omega t,\sin\Theta_0\sin\omega t,\cos\Theta_0), \\
 \end{aligned}
 \, ,
\end{equation}
to express the noise $\mathbf{b}(t)$ in what follows.
We assume $\mathbf{B}(t)$ (and, hence, $\mathbf{b}(t)$) is periodic, and express $\Phi$  in terms of the components of $\mathbf{b}$ in the adapted frame, keeping up to quadratic terms in $\epsilon$, 
\begin{equation}
\label{phiTotal}
\Phi(t)
=
\omega t
+ 
\epsilon  \csc \Theta_0 b_2(t) 
-\epsilon ^2 
( \cot \Theta_0 \csc \Theta_0 b_1(t) b_2(t) + \csc \Theta_0 b_2(t) b_3(t))
\,.
\end{equation}
Note that $\dot{\Phi}(t) >0$ so that the curve $C$ can also be parametrized by $\Phi$. To this end, we invert the above expression, up to $\mathcal{O}(\epsilon)$,
\begin{equation}\label{tFunctionPhi}
t(\Phi)=\frac{1}{\omega}(\Phi - \epsilon \csc\Theta_0 b_2(\Phi/\omega))+\mathcal{O}(\epsilon^2)\,,
\end{equation}
giving for the geometric phase
\begin{equation}
 \varphi_g = -\frac{1}{2}\int_S\sin\Theta d\Theta d\Phi = -\frac{1}{2}\int\limits_0^{2\pi}\{1-\cos\Theta(t(\Phi))\}d\Phi=
 - \pi+\frac{1}{2}\int\limits_0^{2\pi}\cos\Theta(t(\Phi))d\Phi
 \, ,
\end{equation}
where $S$, in the first equality above, denotes the area enclosed by $C$ on the unit sphere.
In terms of the components of $\mathbf{b}$ in the adapted frame, $\cos\Theta$ can be expressed as
\begin{equation}\label{cosThetaTotal}
\cos\Theta(t(\Phi))
=
\cos \Theta_0
- \epsilon  \sin \Theta_0 \tilde{b}_1(\Phi)
+ \epsilon ^2 
\left\{ 
\sin \Theta_0  \tilde{b}_1(\Phi) \tilde{b}_3(\Phi) 
-\frac{1}{2}\cos \Theta_0 
\left(
\tilde{b}_1^2(\Phi)+\tilde{b}_2^2(\Phi)
\right) 
\right\} 
\,, 
\end{equation}
with $\tilde{b}_i(\Phi) \equiv b_i(t(\Phi))$, giving for $\varphi_g$,
\begin{align}
\nonumber
 \varphi_g 
 = &
 \varphi_+ 
 - \frac{\epsilon}{2}\sin\Theta_0\int\limits_0^{2\pi} \tilde{b}_{1}(\Phi)d\Phi 
 \\
 {} &\quad {}
 + \frac{\epsilon^2}{2} 
\int\limits_0^{2\pi}\left\{
  \sin \Theta_0  \tilde{b}_1(\Phi) \tilde{b}_3(\Phi) 
 -\frac{1}{2}\cos \Theta_0 
 \left(
 \tilde{b}_1^2(\Phi)+\tilde{b}_2^2(\Phi)
 \right)  \right\}d\Phi
 \, ,
  \label{geometricPhaseRealization}
\end{align}
with $\varphi_+$ given by (\ref{gamma0}). 
This is the geometric phase associated with a single realization of the stochastic process. It is easily shown, by transforming from our adapted frame to the cartesian one, that the first two terms in the above expression  coincide with the result in Eq.~(14) in Ref. \cite{PhysRevLett.91.090404}.

The final step in our computation consists in performing an average over all possible realizations of the noise, the result depending on the particular statistics assumed. A set of assumptions that we find both minimal and natural is
\begin{equation}
\label{btstat}
\overline{\rule{0ex}{2.4ex} \tilde{b}_i(\Phi)}=0
\, ,
\quad
\overline{\rule{0ex}{2.4ex} \tilde{b}_1(\Phi)\tilde{b}_3(\Phi)}=0
\, ,
\quad
\frac{\partial}{\partial \Phi}
\left(
\overline{\rule{0ex}{2.4ex} {\tilde{b}_1}^2(\Phi)+ {\tilde{b}_2}^2(\Phi)}
\right)
=0
\, ,
\end{equation}
where a bar denotes ensemble average, leading to
\begin{equation}
\label{averageGeometricPhaseStochastic}
\begin{aligned}
 \overline{\rule{0ex}{1.3ex}\varphi}_g 
 &=\varphi_+ -  \frac{\epsilon^2 \pi}{2} \cos\Theta_0
 \left(
 \overline {\rule{0ex}{2.4ex} \tilde{b}_1^2} 
 +  \overline{\rule{0ex}{2.4ex} \tilde{b}_2^2}\right)
 \, .
\end{aligned}
\end{equation}
Our choice to define the statistics of $\tilde{\mathbf{b}}(\Phi)$, rather than those of $\mathbf{b}(t)$ stems from the fact that the geometric phase is independent of the parametrization of the curve $C$, and while $\Phi$ parametrizes $C$ uniquely, the $t$-parametrization is arbitrary. For reference purposes, the geometric phase for a single realization in terms of $t$ rather than $\Phi$ is   
\begin{align*}
\varphi_g 
&=  \varphi_+ 
- \frac{\epsilon\omega}{2}
\sin\Theta_0\!\int\limits_0^T\!\! b_{1}(t)dt 
+
 \epsilon^2
 \!\int\limits_0^T
 \!\!\left[-\frac{\omega}{4}\cos\Theta_0
 \left(b_{1}^2(t) + b_{2}^2(t)\right)
 -\frac{1}{2}b_{1}(t)\dot{b}_{2}(t) 
 \right.
 \\
  & \qquad
 \left.
 + \frac{\omega}{2}\sin\Theta_0 b_{1}(t)b_{3}(t) \right]dt\,
\end{align*}
with $T=2\pi/\omega$.

The main result of this section, equation (\ref{averageGeometricPhaseStochastic}), can be written in a way that resembles (\ref{gamma0sigmaz}). First note that, for a single realization  of the noise, $\cos\Theta=\langle\sigma_z\rangle$ holds, due to the fact that the spin is aligned with $\mathbf{B}(t)$. Starting now from~(\ref{cosThetaTotal}) and using the assumed statistics, (\ref{averageGeometricPhaseStochastic}) can be written as
\begin{equation}\label{meanGPhasesigmaz}
\overline \varphi_g  = -\pi (1-\overline{\langle\sigma_z\rangle})
\,,
\end{equation}
an expression that should be compared to~(\ref{gamma0sigmaz}).  The operational description of how to actually compute the r.h.s.{} of~(\ref{meanGPhasesigmaz}) goes as follows: for a particular realization of the noise $\tilde{\mathbf{b}}(\Phi)$, corresponding to the curve $C$, choose a certain point $\Phi$ along $C$. The spin, at that point, is ``aligned'' with the  total magnetic field $\mathbf{B}(t(\Phi))$. Compute the expectation value of the operator $\sigma_z$ in that state of the spin, and average the result over all possible realizations of the noise --- the  result is $\overline{\langle\sigma_z\rangle}$, which, when plugged in~(\ref{meanGPhasesigmaz}), gives the average phase $\overline \varphi_g$.
\subsection{Comparison with related results in the literature}
\label{Cwrritl}
\noindent We comment here on previous, related work in the literature. A setup similar to ours is considered in Ref. \cite{filipp:2008}, where the noise is taken to have only $z$-component and the analysis is numerical. For each value of $\Theta_0$, four hundred realizations of the noise are generated, the geometric phase for each of them is computed numerically, and its average (over the four hundred runs) is found to deviate from the unperturbed (zero-noise) result. The difference between the two is plotted for various values of $\Theta_0$, and the observation is made that the points obtained are approximated well by a curve proportional to $\sin^2\Theta_0 \cos \Theta_0$.
Using our result, it is not difficult to get to this functional form analytically. First, we note that the appropriate starting point, in this case, is Eq.~(\ref{geometricPhaseRealization}), rather than~(\ref{averageGeometricPhaseStochastic}), since the latter assumes the statistics of~(\ref{btstat}), which, however, are violated by the assumption that the noise only has $z$-component. Indeed, with this latter assumption, we get
\begin{equation}
\label{btstat2}
\overline{\tilde{b}_1\tilde{b}_3}=-\sin\Theta_0 \cos\Theta_0 \, \overline{\tilde{b}_z^2}
\, ,
\qquad
\overline{\tilde{b}_1^2+\tilde{b}_2^2}=\sin^2\Theta_0 \, \overline{\tilde{b}_z^2}
\, ,
\end{equation}
and then~(\ref{geometricPhaseRealization}) gives
\begin{equation}
\label{aGP2}
\Delta \varphi_g 
\equiv 
\varphi_+-\overline{\varphi}_g
=
\frac{\epsilon^2 3\pi \overline{\rule{0ex}{2ex}\,  \tilde{b}_z^2}}{2}  \,  \sin^2\Theta_0\cos\Theta_0
 \, .
 \end{equation} 
Noting than the correspondence between our quantities and those in Ref. \cite{filipp:2008} is $\epsilon \tilde{b}_z \rightarrow 2P/\omega_L$, we rewrite~(\ref{aGP2}) as
\begin{equation}
\label{aGP3}
\Delta \varphi_g 
=
\frac{3\pi}{2} \frac{4P^2}{\omega_L^2} \,  \sin^2\Theta_0\cos\Theta_0
 \, .
 \end{equation} 
By appropriately enlarging Fig.~3(b) of Ref. \cite{filipp:2008}, printing it out, and measuring directly on it with a ruler, we find the maximum value of $\Delta\varphi_g$ to be $1.80 \times 10^{-3}$ rad. On the other hand, the maximum value of $\sin^2\Theta_0 \cos\Theta_0$, obtained for $\Theta_0=.955$, is .385, so that~(\ref{aGP3}) gives (with $\omega_L=3600$ rad/sec)
\begin{equation}
\label{PDeltaphi}
P=1800 \sqrt{\frac{2(\Delta\varphi_g)_\text{max}}{ 0.385 \times 3\pi}}=56.7 \text{rad/sec}
\, .
\end{equation}
This is about half of the value $P=112$ rad/sec that is reported in Ref. \cite{filipp:2008}  --- the discrepancy is far more than can be justified by our approximations. Our best guess as to the origin of this erroneous factor of two is the following: the author of that work reports magnetic fields as frequencies, so that, for example, his $\omega_L$, encoding the strength of the unperturbed magnetic field, is given by $\omega_L=2\mu |B|/\hbar$, the factor of two in this case coming from the fact that $\omega_L$ actually refers to the energy gap between the two spin eigenstates. We suspect the same formula (including the factor of two) was used to convert the noise strength to frequency, resulting in double the correct answer. Assuming that this is indeed so, then our result $P=56.7$ should be compared to the actual value used in the simulations of Ref. \cite{filipp:2008}, $P=56$, with which it is in excellent agreement. Curiously, a least square fit to the data points in Fig.~3(b) of Ref. \cite{filipp:2008}, with $P$ as the sole free parameter, is reported to give $P_\text{fit}= 138.9$ rad/sec, with a reduced $\chi^2$-value of 1.5. Taking into account the above factor of two, this becomes 69.4, which is an unexpected, and unjustified, 24\% off   
the correct value 56. However, the author of Ref. \cite{filipp:2008} uses the incorrect formula $\overline{\varphi}_g=-\pi(1-\cos\bar{\theta})$ for the average geometric phase (just below his Eq.~(18)), instead of the correct one $\overline{\varphi}_g=-\pi(1-\overline{\cos\theta})$. It is easy to show that this results in an erroneous factor of $\sqrt{3/2}\approx 1.22$ for $P_\text{fit}$ --- when this is taken into account (in addition to the above factor of two) one gets $P_\text{fit}=56.7$, in spectacular agreement with our result above. 

Apart from this  numerical treatment, an experimental study of the robustness of the geometric phase has been undertaken  previously.\cite{PhysRevLett.102.030404} The emphasis there though is on the dependence of the spread of the geometric phase (due to noise) on the total evolution time, rather than the shift of its mean value. The single mention of the average phase (below Fig.~4) should not be confused with our use of the term, as the averaging there is performed over different values of the evolution time.
\subsection{The case of non-periodic noise}
\label{Tconpn}
In the preceding analysis the noise $\mathbf{b}(t)$ was assumed periodic, so that the corresponding function $\Theta(t(\Phi))$ was continuous for all $\Phi$ --- here we comment on the case where $\mathbf{b}(t)$ is non-periodic, so that $\Theta(t(0)) \neq \Theta(t(2\pi))$ in general. The geometric phase corresponding to $\Phi$ ranging in the interval $[0,2\pi)$ may still be defined by closing the $\Theta(\Phi)$ curve along a geodesic segment,\cite{PhysRevLett.60.2339} which in this case runs along the meridian $\Phi=0$ --- the phase is still given by minus one-half the area enclosed by the (thus closed) curve. Keeping the same statistics, Eq.~(\ref{btstat}), we are led to identical results, as the periodicity requirement does not enter in the calculation.

An alternative (different) way of treating the problem, is to consider an aperiodic curve as the limiting case of a periodic one, with period $2n\pi$, as the integer $n$ tends to infinity. We can then calculate the geometric phase accumulated in a period (\ie, after $n$ complete turns around the sphere), and compare it to $n$ times that accumulated in the  periodic (period $2\pi$) case --- the two results are again identical.

A third way to address the issue, different from both mentioned above, is to assign to each non-periodic curve a periodic one, by smoothly modifying it in the interval $(2\pi-\delta,2\pi)$, so as to close nicely onto itself. The two curves differ in their enclosed area by an amount of the order of $\epsilon \delta$, since the width of the interval over which they differ is $\delta$ while their vertical separation is of the order $\epsilon$ of the noise amplitude. The minimum value $\delta$ can take is constrained by our perturbative treatment of the problem, in which it was assumed that the $\Phi$-derivative of $\Theta$ was of order $\epsilon$, which implies that the maximum frequency $k_{\text{max}}$ present in $\Theta(\Phi)$ satisfies $k_{\text{max}} \ll \epsilon^{-1}$. The minimum $\delta$ is then of order $k_{\text{max}}^{-1}$, hence the corresponding area difference between the two curves is of order $\epsilon/k_{\text{max}}$. When considering  the various realizations of the noise, this area will average to zero, to the leading order in $\epsilon$. Higher order corrections, due to the asymmetry between the north and south pole  will emerge (when the curve lies away from the equator), so that, after averaging, the area difference will be of order at most $\epsilon^2/k_{\text{max}}$, \ie, suppressed by a factor of $k_{\text{max}}$ w.r.t.{} the quadratic corrections computed above. \eg, for $k_{\text{max}} \sim 10^2$, the two results agree to within one percent. 
\section{Spin 1/2 coupled to a quantum vector operator}
\label{quantumField}
\subsection{Second order corrections to the geometric phase for a quantum driving field}
\label{Socttpfaqdf}
\noindent 
We now replace the classical magnetic field of the previous section by a quantum vector operator $\mathbf{A}$, which corresponds to some physical property of an auxiliary system, \eg, it could be the angular momentum operator of another particle --- its purpose is to provide a ``quantum driving field'' for the spin-1/2, with quantum fluctuations replacing the stochastic classical noise discussed earlier. A useful mental picture derives from a Feynmanesque ``sum over histories'' approach, in which the tip of $\mathbf{A}$ follows all possible paths, each appropriately weighted by a phase factor, the spin-1/2 adiabatically following $\mathbf{A}$ in each of the histories being superposed. Our treatment will be perturbative, corresponding to considering the above multitude of possible paths as mainly that of the expectation value of $\mathbf{A}$, which we take to precess like $\mathbf{B}_0$ did in the previous section, plus a small fluctuating component.

The hamiltonian for the composite spin-particle system is
\begin{equation}\label{totalhamiltonian}
 H = H_A + \frac{\lambda}{2}\, \mathbf{A}\cdot\boldsymbol{\sigma}
 \, ,
\end{equation}
and acts in the tensor product $\mathcal{H}=  \mathcal{H}_\text{particle} \otimes \mathcal{H}_\text{spin}$, with $H_A$, $\mathbf{A}$, acting on $\mathcal{H}_\text{particle}$ and $\boldsymbol{\sigma}$ acting on $\mathcal{H}_\text{spin}$ --- a more explicit notation would be
\begin{equation}\label{totalhamiltonianexp}
 H = H_A \otimes \mathbf{1}_\text{spin} + \frac{\lambda}{2}\, \sum_{i=1}^3 A_i \otimes \sigma_i
 \, .
\end{equation} 
The vector nature of $\mathbf{A}$ is codified in the commutation relations
\begin{equation}
 [L_i,A_j] = i\sum_k \epsilon_{ijk}A_k, \qquad i,j,k = x,y,z,
\end{equation}
between the components of $\mathbf{A}$ and the particle angular momentum components $L_i$. These relations imply that $A_\pm = (A_x\pm iA_y)$ are raising and lowering operators for $L_z$, \ie,
\begin{equation}
 [L_z,A_\pm] = \pm A_\pm
\end{equation}
holds. We assume that $H_A$ commutes with  $\mathcal{A}\cdot\mathbf{L}$, where $\mathcal{A}\equiv\langle n,m|\mathbf{A}|n,m\rangle = \rho\hat{n}$ is the expectation value of $\mathbf{A}$ in the state $|n,m\rangle$, the latter being a simultaneous eigenstate of $H_A$ and $\hat{n} \cdot \mathbf{L}$, with eigenvalues $E_n$  and $m$ respectively. With a perturbative treatment in mind, we write $\mathbf{A}$ in the form
\begin{equation}\label{ARhor0}
 \mathbf{A} = \rho\hat{n} + r_0\, \mathbf{b}
 \,,
\end{equation}
 that is, we decompose $\mathbf{A}$ as a sum of its expectation value in the state $|n,m\rangle$ and its quantum fluctuations, the scale of the latter being carried by $r_0$, so that the expectation value of $b^2 \equiv \mathbf{b} \cdot \mathbf{b}$, in the above state, is of order 1 (the expectation value of $\mathbf{b}$ itself is zero). The total hamiltonian is now expressed as
\begin{equation}\label{hamiltonianAnsatz}
 H = H_A + \frac{\Omega}{2}\hat{n}\cdot\boldsymbol{\sigma} + \epsilon\frac{\Omega}{2}\mathbf{b}\cdot\boldsymbol{\sigma},
\end{equation}
where $\Omega\equiv\lambda\rho$ and $\epsilon\equiv r_0/\rho\ll 1$, the latter inequality deriving from our assumption that the scale of the quantum fluctuations of $\mathbf{A}$ is small compared to that of its expectation value $\mathcal{A}$.

Following Berry's scheme, we assume the system starts in an instantaneous eigenstate of $H$, and the precession of $\mathcal{A}$ is slow enough for it to remain in such a state for all $t$. We compute $H$-eigenstates perturbatively, by writing $H=H_0+\epsilon V$, where
\begin{equation}\label{hamiltonianNonPerturbed}
H_0 = H_A + \frac{\Omega}{2}\sigma_z
\, ,
\qquad
V=\frac{\Omega}{2} \mathbf{b} \cdot \boldsymbol{\sigma}
\, .
\end{equation}
In the above equation, we put momentarily $\hat{n}=\hat{z}$, to facilitate the computation of the eigenfunctions, and we will then rotate the eigenfunction found back 
to a general orientation.

It is clear that $|n,m\rangle\otimes|\hat{z}+\rangle$ is an eigenstate of $H_0$, denoted in the following by $|n,m,+\rangle_0$, with corresponding eigenvalue $E_n+\Omega/2$.
 It is also easy to show that $J_z=L_z + S_z$ commutes with $H$, to all orders in $\epsilon$, where $S_i \equiv  \sigma_i/2$. Since $|n,m,+\rangle_0$ is an eigenstate of $J_z$ with eigenvalue $m+1/2$, it follows by continuity in $\epsilon$ that when the perturbation is turned on, the eigenstate $|n,m,+\rangle$ of the perturbed hamiltonian $H$ will also be a $J_z$ eigenstate, to all orders of $\epsilon$, with the same eigenvalue $m+1/2$. This implies that
\begin{equation}
\label{totalEigenstate}
 \langle\mathbf{x}|n,m,+\rangle 
 = 
 \psi_{n+}(r,\theta)\e^{im\phi}|\hat{z}+\rangle 
 + 
 \psi_{n-}(r,\theta)\e^{i(m+1)\phi}|\hat{z}-\rangle,
\end{equation}
where $\theta$, $\phi$ are the standard polar angles and we have taken into account that the $\phi$-dependence of $L_z$ eigenfunctions, with eigenvalue $m$, is of the form $\e^{im\phi}$. The second term has an extra $\e^{i\phi}$ factor to compensate the spin flip, keeping the $J_z$ eigenvalue equal to $m+1/2$. A rotation of the above wavefunction  in both particle and spin space, bringing $\hat{z}$ to $\hat{n}$, produces a simultaneous eigenstate of $H$ and $\hat{n}\cdot\mathbf{J}$, with eigenvalue $m+1/2$ --- the ambiguity in selecting one of the various rotations that bring $\hat{z}$ to $\hat{n}$ only imparts a phase factor to the resulting eigenfunction.

Assume now that $\hat{n}$ precesses along the circle $(\sin\Theta_0\cos\Phi,\sin\Theta_0\sin\Phi,\cos\Theta_0)$, with $\Phi$ ranging from 0 to $2\pi$. The instantaneous eigenfunction of $H$ is obtained by that given in~(\ref{totalEigenstate}) by an appropriate rotation. This is a $\hat{n} \cdot \mathbf{J}$, $m+1/2$ eigenstate, so that the geometric phase it accumulates is
\begin{equation}
 \varphi =-(2m+1)\pi(1-\cos\Theta_0)
 \, .
\end{equation}
The above is the geometric phase accumulated by the entire system, particle plus spin.
Having in mind the eventual comparison with the ``classical'' case of the previous section, we now wish to subdivide this total phase to the two subsystems, the particle and the spin. How exactly to accomplish this has been the subject of a certain dispute --- see, \eg, an original proposal in Ref. \cite{PhysRevLett.92.150406}, its subsequent questioning in Ref. \cite{PhysRevLett.100.168901}, and the reply in Ref. \cite{PhysRevLett.100.168902}. Our understanding of the matter is that a proper definition of this subdivision must be tied to a particular operational recipe for its measurement. In this sense, we consider the question still open in principle, both definitions currently available in the literature being applicable to particular situations. For our purposes here, we adopt the definition in Refs. \cite{Tong_2003,PhysRevLett.92.150406,PhysRevLett.99.139903}, based on its intuitive appeal, as well as on the fact that it leads to results compatible with the classical ones of the previous section --- the ultimate answer on which is the right way to proceed can only be decided experimentally. 

We start with the  Schmidt decomposition~\cite{peres:1995} of the perturbed eigenstate
$ |n,m,+\rangle$,
\begin{equation}
\label{ketnm+}
 |n,m,+\rangle 
 = 
 \sqrt{p_+} \, |e_+\rangle \otimes | +\rangle 
 + 
 \sqrt{p_-} \, |e_-\rangle \otimes |-\rangle
 \, ,
\end{equation}
where $|e_\pm\rangle$, $|\pm\rangle$ are orthonormal states in $\mathcal{H}_\text{particle}$, $\mathcal{H}_\text{spin}$, respectively, and $p_+ + p_-=1$. Then, the spin geometric phase $\varphi_\text{spin}$ is defined as the weighted average
\begin{equation}
 \varphi_\text{spin} = p_+\varphi_+ + p_-\varphi_-
 \, ,
\end{equation}
where $\varphi_\pm$ are the standard geometrical phases that would be accumulated by the spin if it was aligned or antialigned, respectively, with $\hat{n}$ during its precession.

Inspection of the wavefunction in~(\ref{totalEigenstate}) is enough to convince that it is already in the Schmidt form, with 
\begin{equation}
\label{psiSchmidt}
p_\pm = 2\pi \int_0^\infty dr \, r^2 \int_0^\pi d\theta \, 
\sin \theta  |\psi_{n\pm}(r,\theta)|^2
\, .
\end{equation}
With $\varphi_\pm = \mp\pi(\cos\Theta_0-1)$, we get
\begin{equation}
\label{spinPhaseSchmidt}
 \varphi_\text{spin} = -\pi(1-\cos\Theta_0)p_+ +\pi(1-\cos\Theta_0)p_- = -\pi(1-\cos\Theta_0) - 2\pi\cos\Theta_0 \,  p_-
 \, .
\end{equation}
Using the fact that $\ket{{\pm}}$ are eigenstates of $\boldsymbol{\sigma}\cdot \hat{n}$, and equation (\ref{ketnm+}), we can write the previous result in the form
\begin{equation}
\varphi_\text{spin}  = -\pi(1-\langle n,m,+|\sigma_z|n, m, +\rangle)
\, ,
\end{equation}
comparison of which with those in~(\ref{gamma0sigmaz}) and~(\ref{meanGPhasesigmaz}) further justifies our procedure for assigning a geometrical phase to the spin subsystem. Note that the first term in the r.h.s.{} of~(\ref{spinPhaseSchmidt}) is the standard ``classical'' phase, while the second one represents quantum corrections due to the uncertainty in the direction of $\mathbf{A}$. The latter results in a nonzero probability $p_-$ for the spin to antialign, upon measurement, to $\mathcal{A}$, and the quantum correction to the phase is proportional to this spin flip probability, which we now proceed to compute.

The \emph{unnormalized} perturbed state of the composite system can be formally written as $\widetilde{|n,m,+\rangle} = |n,m,+\rangle_0 + \epsilon |n,m,+\rangle_1 + \epsilon^2 |n,m,+\rangle_2$, keeping up to quadratic terms in $\epsilon$, so that, after normalization, we have 
\begin{equation}
 |n,m,+\rangle 
 = 
 \left(
 1-\frac{\epsilon^2}{2} {}_1\langle n,m,+|n,m,+\rangle_1
 \right)
 \left(
 |n,m,+\rangle_0 
 +
  \epsilon |n,m,+\rangle_1 + \epsilon^2 |n,m,+\rangle_2
  \right)
  \, .
\end{equation}
Denoting by $|n,m,+\rangle_i^\pm$ the spin up/down component of $|n,m,+\rangle_i$ and recalling that $|n,m,+\rangle_0$ has no spin down component, the spin flip probability turns out to be 
\begin{equation}
\label{pmref1}
p_- = \epsilon^2 \, \, {}_1^{\! \! -} \! \langle n,m,+|n,m,+\rangle_1^-
\, .
\end{equation} 
Using first order time-independent perturbation theory, we get
\begin{equation}\label{firstorderperturbation}
 \begin{split}
 |n,m,+\rangle_1^- & 
 = 
 \sum_k \frac{1}{E_{nk}+\Omega}
 |k,m,-\rangle_0\,{}_0\langle k,m,-|V|n,m,+\rangle_0 
\\
 & = 
\frac{\Omega}{2}\sum_k \frac{1}{E_{nk}+\Omega}
|k,m,-\rangle_0\,{}_0\langle k,m,-|\mathbf{b}\cdot\boldsymbol{\sigma}|n,m,+\rangle_0 
\\
& = 
\frac{\Omega}{2}\sum_k \frac{1}{E_{nk}+\Omega}
|k,m\rangle|\hat{n}-\rangle\langle k,m|b_+|n,m\rangle 
\\
\end{split}
\end{equation}
where $b_\pm =b_1\pm ib_2$ and $E_{nk} = E_n-E_k$. The last expression simplifies in the adiabatic limit we are considering. Indeed,  starting from Heisenberg's equation of motion for an operator $Q$, $\dot{Q}_t = i[H,Q_t]$, we find that the matrix elements of $Q_t$ in an energy basis satisfy the equation $\dot{Q}_{nm} = E_{nm}Q_{nm}$ (no summation involved), so that the frequency spectrum of $Q_t$ consists in the energy differences of the hamiltonian eigenstates it connects. In the adiabatic approximation, the time evolution of $\mathbf{b}_t$ is much slower than $\Omega$, so that the frequencies it contains in its spectrum can be neglected in comparison with $\Omega$. Therefore, in the last expression above, the denominator becomes just $\Omega$ and the sum over $k$ is the identity in $\mathcal{H}_\text{particle}$, yielding $|n,m,+\rangle_1^- = b_+|n,m,-\rangle_0/2$ and therefore
\begin{equation}\label{quantumCorrections}
 p_- = \frac{\epsilon^2}{4}\langle n,m|b_- b_+|n,m\rangle = \frac{\epsilon^2}{4}\langle n,m|b_1^2 + b_2^2 + i[b_1,b_2]|n,m\rangle.
\end{equation}
The geometric phase for the spin is, from (\ref{spinPhaseSchmidt}),
\begin{equation}\label{quantumPhase}
 \varphi_\text{spin} = -\pi(1-\cos\Theta_0) - \frac{\epsilon^2 \pi}{2}\cos\Theta_0\langle n,m|b_1^2 + b_2^2 + i[b_1,b_2]|n,m\rangle.
\end{equation}
The first term is the classical geometric phase associated with the precession of $\mathcal{A}$. The second term is the quadratic quantum correction we have been after.  Note that the  two terms $b_1^2 + b_2^2$ and $[b_1,b_2]$ are individually invariant under rotations in the 1-2 plane. The first of these is proportional to  the solid angle subtended by the fluctuations of $\mathbf{A}$, as seen from the origin. Comparison of this term, with the corresponding one in the r.h.s.{} of~(\ref{averageGeometricPhaseStochastic}) suggests a close analogy, but a careful analysis reveals subtleties that need to be adressed before claiming ``agreement''.
\subsection{Correspondence between the stochastic and quantum cases}
\noindent 
Classical and quantum fluctuations, as considered in this work, differ in a fundamental way, that renders their comparison nontrivial: on the one hand, classical noise involves stochastic processes, each realization of which is an arbitary function of time, and the statistical properties of which can be assigned at will. On the other hand, quantum noise is modelled here by a vector operator, which corresponds to a physical quantity, and has its own dynamics, generated by a hamiltonian, while expectation values are with respect to a particular eigenstate of the hamiltonian --- it is clear that  the quantum case is much more constrained than the classical one and that, therefore, the two approaches may lead to analogous results only if care is taken to carefully match the corresponding systems. To this end, first we define the correspondence between classical and quantum quantities, 
\begin{equation}
\tilde{b}_i \rightarrow b_i
\, ,
\qquad
\tilde{b}_1 \tilde{b}_3 \rightarrow \frac{1}{2}(b_1 b_3+b_3 b_1)
\, ,
\end{equation}
the visual simplicity of which is somewhat misleading, as the tilded classical functions of $\Phi$ are those that are mapped to operators. We gave already a plausibility argument  
for working with the functions $\tilde{b}_i(\Phi)$, instead of the $b_i(t)$, but we feel that a deeper, more satisfactory explanation of why ``things work'' with this choice is due. 
Second, we replace ensemble averages with quantum expectation values in a particular quantum state, $\overline{\rule{.3ex}{0ex}\cdot \rule{.3ex}{0ex}} \rightarrow \langle n,m|\cdot |n,m\rangle$.
Then we show in what sense are conditions analogous to those in~(\ref{btstat}) satisfied in the quantum case. The first of~(\ref{btstat}) is mapped, as defined above,  to the relation $\langle n,m| b_i|n,m\rangle=0$, which is satisfied by definition, as the expectation value of $\mathbf{A}$ is carried in its entirety by $\mathcal{A}$. The second of~(\ref{btstat}) is mapped to the relation 
$\langle n,m|\, b_1 b_3+b_3 b_1 \, |n,m\rangle=0$, which can easily be shown to hold, given the vector nature of $\mathbf{A}$ and the fact that the state $|n,m\rangle$ is an $L_3$-eigenstate. Finally, the third of~(\ref{btstat}) is mapped to the relation $\partial/\partial \Phi \langle n,m| b_1^2+b_2^2 |n,m\rangle=0$, which  follows at once from symmetry considerations. Note a certain abuse of notation in all of the above relations, in that $|n,m\rangle$ actually refers to the $H$-eigenstate for a general $\hat{n}$ (\ie, for a general $\Phi$), but as the frame used rotates the same way the wavefunction does, with varying $\Phi$, all expectation values are $\Phi$-independent, and that is why we have suppressed $\Phi$ in our notation.

From the previous observations, by replacing the ensemble averages with expectation 
values in (\ref{averageGeometricPhaseStochastic}), we obtain the following expression for the geometric phase: 
\begin{equation}\label{GeometricPhaseQuantic}
 \varphi_\text{spin} =  
 \varphi_+ - \frac{\epsilon^2 \pi}{2}\cos\Theta_0  
 \langle n,m| b_1^2 + b_2^2 |n,m \rangle 
 \, ,
\end{equation}
which is the same as (\ref{quantumPhase}), except, importantly, for the commutator term, which has no classical analogue. 
\subsection{Examples}
\noindent We apply our main result, Eq.~(\ref{quantumPhase}), to some physical systems, to illustrate its implications, and, in particular, the crucial role of the commutator term in the consistency of the whole scheme.
\subsubsection{The SHO case}
\noindent Consider the case where $\mathbf{A}$ is the position operator of a three dimensional harmonic oscillator. This is a ``mildly'' quantum case, as  the components $A_i$ commute among themselves. We accomplish the precession of the mean position of the oscillator by positioning the minimum of the parabolic potential of the oscillator off the $z$-axis, and then rotating it adiabatically around the $z$-axis, so that the particle follows the potential, in a fixed quantum state, that we choose to be the ground state $|0\rangle$. We can, alternatively, position the minimum of the potential on the $z$-axis, at some positive $z_0$, and consider a coherent state that rotates in the $z=z_0$ plane.
In this case, the commutator term in~(\ref{quantumPhase}) is zero,  while 
$\langle 0|b_1^2+b_2^2|0\rangle = 2$, so that
\begin{equation}
\begin{split}
\varphi_s & = -\pi(1-\cos\Theta_0)-\pi\epsilon^2\cos\Theta_0 \\
	            & \equiv -\pi(1-\cos\Theta_\text{eff}) 
	            \, 
	            \end{split}
	\end{equation}
	with $\Theta_\text{eff} = \Theta_0 + \epsilon^2\cot\Theta_0 + \mathcal O(\epsilon^4)$ an effective polar angle. For $\Theta_0 = \pi/4$ and $\epsilon = 0.05$, $\Theta_\text{eff}$ is about 0.14 degrees larger than $\Theta_0$, signifying a slight increase (about 1\%) in the geometric phase accumulated in one period.
\subsubsection{The spin $\frac{1}{2}$ case}
\noindent 
We now consider a ``fully quantum'' case where the components $A_i$ of  $\mathbf{A}$ do not commute. Our starting point is the hamiltonian
\begin{equation} 
H = \mathbf{B}\cdot\mathbf{s}_1 + \lambda \, \mathbf{s}_1\cdot\mathbf{s}_2
\, ,
\end{equation}
describing a spin-1/2 $\mathbf{s}_1$ coupled to an adiabatically precessing magnetic field $\mathbf{B}$ and a second spin-1/2 $\mathbf{s}_2$, coupled to $\mathbf{s}_1$. The field direction $\hat{m}$ precesses around the $\hat{z}$-axis at some fixed angle $\Theta_0$, while  $\mathbf{s}_1$ remains in the instantaneous eigenstate $|\hat{m}(t)+\rangle$ --- for this to happen, approximately, $\lambda \ll B$ must hold. Here, the operator $\mathbf{A}$ of our general treatment above is identified with $\mathbf{s}_1$. The astute reader might object that in our treatment the ``particle'' wavefunction was defined over physical space, while here $\mathbf{s}_1$ only has internal degrees of freedom, it is an easy exercise though to retrace our steps, effecting the appropriate modifications, in order to arrive at~(\ref{quantumPhase}) without any restriction on the nature of $\mathbf{b}$.   We compute 
$\mathcal{A}(t) \equiv \rho \hat{m}(t) 
= 
\langle\hat{m}(t)+| \mathbf{s}_1 |\hat{m}(t)+\rangle 
= 
\hat{m}(t)/2$, that is, $\rho=1/2$. The quantum correction to the phase is proportional to \begin{equation}
\label{qcs}
\langle\hat{m}+|\sigma_1^2 + \sigma_2^2 + i[\sigma_1,\sigma_2]|\hat{m}+\rangle 
= 
\langle\hat{m}+|1 + 1 -2\sigma_3|\hat{m}+\rangle = 0
\, .
\end{equation} 
This null result might come as a surprise, as the ``driving field'' $\mathbf{s}_1$ is far from classicality, the uncertainty in its direction being of order one. A simple argument can be given to account for this. An eigenstate of $H$ corresponds to \emph{both} spins aligned with the instantaneous magnetic field $\mathbf{B}(t)$. Choosing this as instantaneous eigenstate, the composite system is then in a state with total spin $1$, following the magnetic field along a circle, and therefore accumulating a geometric phase twice that of a spin 1/2, with either spin accumulating half of the total phase. The implication is that no quantum corrections appear, in all orders in $\epsilon$. It is reassuring that our result~($\ref{quantumPhase}$) is consistent with the above conclusion, with the commutator term being essential for this. The geometric phase for this (factorizable) composite state, lends itself to a simple subdivision among subsystems, coinciding with the treatment presented, in a different setting, previously. \cite{PhysRevA.62.022109}
\subsubsection{The general angular momentum case}
\noindent 
As another example, we consider the hamiltonian
\begin{equation}
H = \mathbf{B}\cdot\mathbf{L} + \lambda \, \mathbf{L}\cdot\mathbf{s}_2,
\end{equation}
where the orbital angular momentum $\mathbf{L}$ of a particle replaces, as ``quantum driving field'', the spin-1/2 $\mathbf{s}_1$ of the previous example.  We consider the particle to be in the $|l,m\rangle$ eigenstate, with $m>0$ labeling eigenstates of angular momentum in the direction $\hat{n}$ of $\mathbf{B}$, so that $\mathbf{L}$ is not necessarily fully aligned with the magnetic field. Then $\langle l,m|\mathbf{L}|l,m\rangle =  m  \hat{n}$ and $\mathbf{s}_2$ aligns with $\hat{n}$. Since $\langle l,m|L^2-L_3^2|l,m\rangle = l(l+1)-m^2$, we get for the geometric phase~(\ref{quantumPhase})
    \begin{equation}\label{amqfield}
    \varphi_s=-\pi(1-\cos\Theta_0)-\pi\cos\Theta_0\left(\frac{l(l+1)-m(m+1)}{2m^2}\right)\,.
    \end{equation}
 Note that quantum corrections to the geometric phase accumulated by $\mathbf{s}_2$ vanish for the highest $\mathbf{L}$-projection state $m=l$ (a spin-coherent state \cite{Bengtsson2008}) --- in this sense, spin-coherent states have classical-like attributes.
\subsubsection{An exact calculation of the quantum correction}
\noindent 
Interestingly enough, a case slightly more general than the previous one can be solved exactly, in the adiabatic limit, without using perturbation theory. We assume that $H_A$ commutes with $L^2$, as well as with $\hat{n} \cdot \mathbf{L}$, and also that the energy eigenvalues $E_{lm}$, with $H|l,m\rangle=E_{lm}|l,m\rangle$ only depend on $l$, $m$. Taking, as before, $\mathbf{A} \rightarrow \mathbf{L}$, and with $H$ as in~(\ref{totalhamiltonian}), we find
 \begin{align}
 H \ket{l,m,+}&=\left(E_{lm}+\frac{\lambda m}{2}\right)\ket{l,m,+}+\frac{\lambda}{2}\sqrt{l(l+1)-m(m+1)}\ket{l,m+1,-}\,,\\
 H \ket{l,m+1,-}&=\left(E_{lm+1}-\frac{\lambda (m+1)}{2}\right)\ket{l,m+1,-}+\frac{\lambda}{2}\sqrt{l(l+1)-m(m+1)}\ket{l,m,+}\,.
 \end{align}
These relations suffice, after some algebra, to diagonalize $H$ and obtain the eigenstate in (\ref{ketnm+}), from which we can read off
 \begin{equation}
 p_-=\frac{l-m}{2l + 1}+\mathcal O\left(\frac{E_{lm+1}-E_{lm}}{\lambda m}\right)\,,
 \end{equation}
with $\ket{e_+}=\ket{l,m}$, $\ket{e_-}=\ket{l,m+1}$ and $p_+=1-p_-$. As mentioned in the explanation following (\ref{firstorderperturbation}), the quantity $(E_{lm+1}-E_{lm})/{\lambda m}$ tends to zero in the adiabatic limit (in this case we have $\Omega=\lambda\rho=\lambda m$), so that using (\ref{spinPhaseSchmidt}) we obtain the following for the geometric phase of the spin subsystem 
\begin{equation}
\varphi_\text{spin}^{\text{exact}}=-\pi(1-\cos\Theta_0)-\pi\cos\Theta_0\left(\frac{2l-2m}{2l + 1}\right)\,,
\end{equation}
while the perturbative  result is (\ref{amqfield}),
which is, in general, different. Note however that, in order for  $\epsilon\ll1$ to hold, so that perturbation theory is valid, we need $l$ to be big, and $m$ to 
be close to $l$ (in this case we have $\epsilon=\sqrt{l(l+1)-m(m+1)}/m$). Taking this into consideration by writing $l=m+k$ with $0 \leq k\ll m$, we obtain that
\begin{equation}
\varphi_\text{spin}^{\text{exact}}=-\pi(1-\cos\Theta_0)-\pi\cos\Theta_0\left(\frac{k}{m}\right)+\mathcal O(m^{-2})=
\varphi_\text{spin}+\mathcal O(m^{-2})\,,
\end{equation}
so that the two results coincide in this limit, as expected.
\section{Concluding remarks}
\label{conclusions}
\noindent 
We analyzed the effect of noise to the average geometric phase accumulated by a spin-1/2 particle coupled adiabatically to a precessing magnetic field, up to terms quadratic in the noise amplitude. Our result shows excellent agreement with numerical simulations reported in Ref. \cite{filipp:2008}. We then studied the geometric phase picked up by a spin-1/2  coupled to a quantum vector operator, our main result being Eq.~(\ref{quantumPhase}). An immediate complication implied by the very nature of the question is the division of the total geometric phase of the system into its subsystems. Although we feel that a complete answer to this question is still lacking, following a natural preexisting proposal leads to a simple result, that recovers, in a certain, precise sense, the classical one, while exhibiting a purely quantum novelty, shown to be necessary on grounds of internal consistency.

Our next step along the above line of research is the generalization of our result to the
higher-spin case and, most importantly, to the Wilczek-Zee effect. This latter direction is of interest to quantum computation, where anholonomies are engineered
to represent logical gates \textemdash{}  quantum corrections to these are relevant to both gate
design and robustness assessment. Another question we are currently investigating
is the relation of the quantum corrections, when $\mathbf{A}$ is an angular momentum operator, to the
Wehrl entropy of the states considered, our motivation stemming from the observation that the spin-coherent states that minimize Wehrl's entropy also nullify the
quantum corrections. Exploring these issues would also suggest approaches to hybrid systems, in particular it should shed light on the classical-to-hybrid-to-quantum
transition.

\section*{Acknowledgments}

The authors wish to acknowledge partial financial support from the PAPIIT program of DGAPA-UNAM, grants IN114712 and IG100316. 


%
\end{document}